# The Impact of Class Rebalancing Techniques on the Performance and Interpretation of Defect Prediction Models

Chakkrit Tantithamthavorn, *Member, IEEE,* Ahmed E. Hassan, *Member, IEEE,* and Kenichi Matsumoto, *Senior Member, IEEE*

**Abstract**—Defect prediction models that are trained on class imbalanced datasets (i.e., the proportion of defective and clean modules is not equally represented) are highly susceptible to produce inaccurate prediction models. Prior research compares the impact of class rebalancing techniques on the performance of defect prediction models. Prior research efforts arrive at contradictory conclusions due to the use of different choice of datasets, classification techniques, and performance measures. Such contradictory conclusions make it hard to derive practical guidelines for whether class rebalancing techniques should be applied in the context of defect prediction models. In this paper, we investigate the impact of 4 popularly-used class rebalancing techniques on 10 commonly-used performance measures and the interpretation of defect prediction models. We also construct statistical models to better understand in which experimental design settings that class rebalancing techniques are beneficial for defect prediction models. Through a case study of 101 datasets that span across proprietary and open-source systems, we recommend that class rebalancing techniques are necessary when quality assurance teams wish to increase the completeness of identifying software defects (i.e., Recall). However, class rebalancing techniques should be avoided when interpreting defect prediction models. We also find that class rebalancing techniques do not impact the AUC measure. Hence, AUC should be used as a standard measure when comparing defect prediction models.

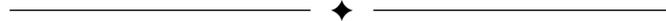

## 1 INTRODUCTION

DEFECT prediction models play a critical role in the prioritization of SQA effort. Defect prediction models are trained using historical data to identify defect-prone software modules. From an SQA perspective, defect prediction models serve two main purposes. First, defect prediction models can be used to predict modules that are likely to be defect-prone in the future [2, 13, 24, 38, 52–54, 62, 96]. SQA teams can use defect prediction models in a prediction setting to effectively allocate their limited resources to the modules that are most likely to be defective. Second, defect prediction models can be used to understand the impact of various software metrics on the defect-proneness of a module [7, 46, 51, 52, 73, 74]. For example, one builds a prediction model using a code complexity metric with an assumption that more complex code shares an increasing relationship with defect-proneness. If the model shows that code complexity is the most important metric (i.e, top-rank metric). Such insights that are derived from defect prediction models can help software teams avoid past pitfalls that are associated with defective modules (e.g., developers should initiate a quality improvement plan that code should be less complex for future releases).

The performance and interpretation of a defect prediction model depend heavily on the data on which it was trained. Prior work raised concerns that defect prediction models that are trained on *imbalanced datasets* (i.e., datasets where the proportion of defective and clean modules is not equally represented) are highly susceptible to producing inaccurate prediction models [26]. Indeed, when training a defect prediction model from an imbalanced dataset, traditional classification techniques often fail to accurately identify the minority class (i.e., defective modules).

To mitigate the risk of imbalanced datasets, prior studies apply *class rebalancing techniques* (i.e., techniques for rebalancing the proportion of defective and clean modules of the training corpus). Such techniques aim to produce an equal representation of two classes of software modules (i.e., defective and clean modules) prior to constructing defect prediction models. Plenty of prior studies have shown a performance improvement when applying class rebalancing techniques in machine learning area. For example, Chawla [8] and Seiffert et al. [70] show that the AUC performance can be substantially improved by up to 40% when applying class rebalancing techniques.

Recent defect prediction studies have compared the impact of class rebalancing techniques on the perfor-

- C. Tantithamthavorn is with the School of Computer Science, the University of Adelaide, Australia.
  E-mail: chakkrit.tantithamthavorn@adelaide.edu.au.
- A. E. Hassan is with the School of Computing, Queen's University, Canada. E-mail: ahmed@cs.queensu.ca.
- K. Matsumoto is with the Graduate School of Information Science, Nara Institute of Science and Technology, Japan. E-mail: matumoto@is.naist.jp.

mance of defect prediction models [34, 45, 63, 65, 68, 78, 90]. For example, Kamei *et al.* [34] show the performance improvement of class rebalancing techniques on 2 defect datasets of proprietary systems. Recently, Malhotra *et al.* [45] show the impact of class rebalancing techniques on 6 defect datasets of open-source systems. In contrast, Riquelme *et al.* [63] argue that class rebalancing techniques have little impact on the performance of defect prediction models when they are trained on 4 datasets of NASA systems. Such contradictory conclusions make it hard to derive practical guidelines about whether class rebalancing techniques should be applied in the context of defect prediction models. Since prior work focuses on the different choice of datasets, classification techniques, and performance measures, it is likely that class rebalancing techniques may be useful for some specific contexts of defect prediction models.

Moreover, Turhan [89] points out that applying class rebalancing techniques may lead to bias in learned concepts (i.e., *concept drift*) — the resampled training dataset is not representative of the original dataset. Indeed, *concept drift* appears when the class distributions of training and testing datasets are different. Thus, class rebalancing techniques may impact the interpretation of defect prediction models.

In this paper, we set out to investigate the impact of 4 popularly-used class rebalancing techniques, i.e., oversampling, under-sampling, SMOTE, and ROSE techniques, on the performance and interpretation of defect prediction models. We train our defect prediction models using 7 commonly-used classification techniques, i.e., random forest (RF), logistic regression (LR), naive bayes (NB), neural network (AVNNet), C5.0 Boosting (C5.0), extreme gradient boosting (xGBTree), and gradient boosting method (GBM). We evaluate the performance of defect prediction models using 10 commonly-used performance measures, i.e., 3 threshold-independent (e.g., AUC) and 7 threshold-dependent (e.g., Precision, Recall, F-Measure) performance measures.

To better understand the impact of class rebalancing techniques on defect prediction models, we construct statistical models to study the relationship between the experimental factors (e.g., defective ratios, classification techniques) and the performance and interpretation of defect prediction models. Through a large-scale empirical study of 101 publicly-available defect datasets that span across open-source and proprietary systems that are collected from 5 different corpus, we address the following research questions:

**(RQ1)** **How imbalanced are defect prediction datasets?**
As little as 8% of defect datasets have a defective ratio between 45%-55%, suggesting that most defect datasets are unbalanced.

**(RQ2)** **How do class rebalancing techniques impact the performance of defect prediction models?**
*Performance Analysis.* The AUC measure is not impacted by class rebalancing techniques for defect prediction models, unlike common findings in machine learning literature where the AUC performance is substantially improved with the applications of class rebalancing techniques. Class rebalancing techniques impact Recall the most positively and impact Precision the most negatively.
*Experimental Factors Analysis.* Class rebalancing techniques yield the largest performance improvement for defect prediction models when applying the under-sampling rebalancing technique to logistic regression models using defect datasets that are highly-imbalanced with an EPV ratio higher than 40.

**(RQ3)** **How do class rebalancing techniques impact the interpretation of defect prediction models?**
*Interpretation Analysis.* Class rebalancing techniques shift the learned concepts (i.e., biasing the interpretation of defect prediction models). We find that as little as 23%-34%, 55%-62%, and 68%-71% of the top variables in the top importance rank of the re-balanced models appear in the top importance rank of the baseline models for neural network, logistic regression, and random forest classifiers, respectively.
*Experimental Factors Analysis.* The impact of class rebalancing techniques on the interpretation of defect prediction models relies heavily on the used classification techniques, suggesting that researchers and practitioners should avoid when deriving knowledge and understandings from defect prediction models.

Our results lead us to conclude that the impact of class rebalancing techniques on the performance of defect prediction models depend on the used performance measure and the used classification techniques. While class rebalancing techniques substantially improve the Recall measure and decrease the Precision measure, they do not impact the AUC measure. On the other hand, class rebalancing techniques negatively impact the interpretation of defect prediction models—i.e., we find that class rebalancing techniques shift the learned concepts to the interpretation of defect prediction models. Based on our findings, we recommend that class rebalancing techniques are beneficial when quality assurance teams wish to increase the completeness of identifying software defects (i.e., Recall), but they should be avoided when deriving knowledge and understandings from defect prediction models. We also find that class rebalancing techniques do not impact the AUC measure. Hence, AUC should be used as a standard measure when comparing defect prediction models.

## 1.1 Novelty Statements

This paper presents the first empirical study to investigate (1) the impact of class rebalancing on defect prediction models using the largest number of commonly-used defect datasets (i.e., 101 defect datasets)—prior studies focus on less than 10 defect datasets; (2) the impact of class rebalancing techniques on the interpretation of defect prediction models; and (3) the experimental design settings where class rebalancing yields the largest benefits for defect prediction models.

## 1.2 Contributions

The contributions of our paper are as follows:

1) An empirical demonstration of the nature of class imbalance in 101 publicly-available defect datasets.

2) An empirical investigation of the impact of class rebalancing techniques on 10 commonly-used threshold-dependent and threshold-independent performance measures.

3) An empirical investigation of the impact of class rebalancing techniques on the interpretation of defect prediction models.

4) An in-depth examination of the impact of experimental factors (including class rebalancing techniques) on the performance and interpretation of defect prediction models.

## 1.3 Paper organization

The remainder of this paper is organized as follows. Section 2 illustrates the nature of class imbalance in defect datasets. Section 3 introduces class rebalancing techniques. Section 4 positions this paper with respect to the related work. Section 5 discusses the design of our case study, while Section 6 presents the results with respect to our two research questions. Section 7 offers practical guidelines for practitioners and researchers. Section 8 discloses the threats to the validity of our study. Finally, Section 9 draws conclusions.

## 2 THE NATURE OF IMBALANCED DEFECT DATASETS

**Motivation**. Class imbalance refers to a classification problem where the classes (i.e., the proportion of defective and clean modules) are not represented equally. However, little is known about the nature of class imbalance in defect prediction datasets. Thus, we set out to investigate the following research question.

> *(RQ1) How imbalanced are defect prediction datasets?*

**Approach**. In order to assess whether class imbalance is prominent in defect prediction studies, we analyze the defective ratio of 101 publicly-available defect datasets that have been popularly studied in prior defect prediction research. 76 datasets are downloaded from the Tera-PROMISE repository [48], 12 clean NASA datasets are provided by Shepperd *et al.* [71], 5 datasets are provided by Kim *et al.* [37] and Wu *et al.* [91], 5 datasets are provided by D'Ambros *et al.* [13, 14], and 3 datasets are provided by Zimmermann *et al.* [96]. Figure 1 shows

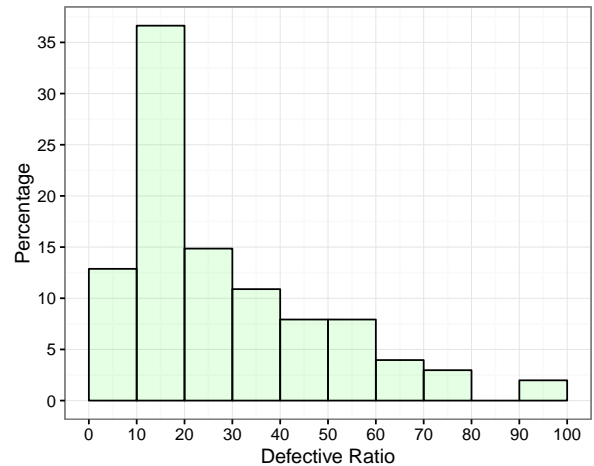

Fig. 1: A histogram of the defective ratios of the 101 publicly-available defect datasets.

a histogram of the defective ratios of the 101 defect datasets.

**Results**. **64% of the defect datasets have a defective ratio below 30%.** Indeed, 38% of the defect datasets have a defective ratio between 10%-20%, suggesting that the majority of defect datasets are highly imbalanced. However, as little as 8% of defect datasets have a defective ratio between 45%-55%, suggesting that there are only few defect datasets that have a defective ratio of nearly 50% (i.e., balanced datasets). On the other hand, only 1% of defect datasets (i.e., `log4j-1.2`,`xalan-2.7`) have a defective ratio higher than 90%.

> *The majority of defect datasets (64%) that are popularly-used in the literature have a defective ratio below 30%, suggesting that class imbalance is prominent in defect datasets, likely affecting the performance and interpretation of defect prediction models.*

## 3 CLASS REBALANCING TECHNIQUES FOR DEFECT PREDICTION MODELS

There are a plethora of class rebalancing techniques available [27], e.g., (1) sampling methods for imbalanced learning, (2) cost-sensitive methods for imbalanced learning, (3) kernel-based methods for imbalanced learning, and (4) active learning for imbalanced learning. Since it is impractical to study all of these techniques, we would like to select a manageable set of class rebalancing techniques for our study. As discussed by He *et al.* [27], we start from the four families of imbalance learning. Based on a literature surveys by Hall *et al.* [20], Shihab [72], and Nam [55], we then select only the family of sampling methods for the context of defect prediction.

We first select the three commonly-used techniques (i.e., over-sampling, under-sampling, and SMOTE [9]) that were previously used in the literature [34, 36, 58, 69, 78, 82, 90, 92–94]. Recent research shows that bootstrap resampling techniques tend to produce more accurate and reliable estimates in the context of software engineering [85]. Recently, Menardi *et al.* [47] show that a smoothed bootstrap resampling technique (ROSE) outperforms other techniques in a non-software engineering domain. Thus, we select the ROSE technique [42, 47] in our study. Figure 2 provides an illustrative overview of the 4 studied class rebalancing techniques. Below, we provide a description and a discussion of the 4 studied class rebalancing techniques for our study.

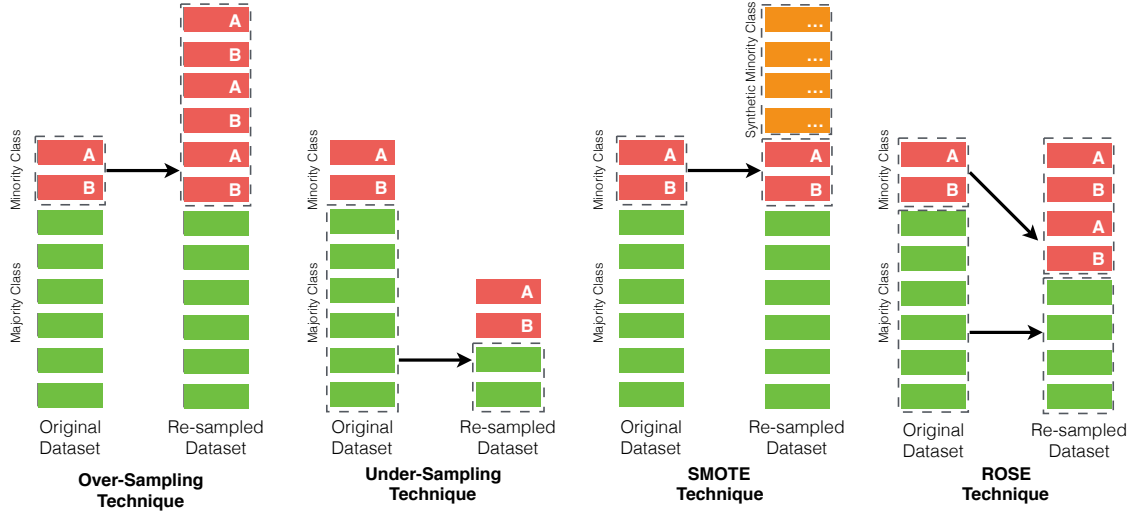

Fig. 2: An illustrative overview of the 4 studied class rebalancing techniques.

### 3.1 Over-Sampling Technique (OVER)

The over-sampling technique (a.k.a. up-sampling) randomly samples with replacement (i.e., replicating) *the minority class* (e.g., defective class) to be the same size as the majority class (e.g., clean class). The advantage of an over-sampling technique is that it leads to no information loss. Since oversampling simply adds replicated modules from the original dataset, the disadvantage is that the training dataset ends up with multiple redundant modules, leading to an overfitting. Thus, when applying the over-sampling technique, the performance of with-in defect prediction models is likely higher than the performance of cross-project defect prediction models.

### 3.2 Under-Sampling Technique (UNDER)

The under-sampling technique (a.k.a. down-sampling) randomly samples (i.e., reducing) *the majority class* (e.g., clean class) in order to reduce the number of majority modules to be the same number as the minority class (e.g., defective class). The advantage of an under-sampling technique is that it reduces the size of the training data when the original data is relatively large. However, the disadvantage is that removing modules may cause the training data to lose important information pertaining to the majority class.

### 3.3 Synthetic Minority Oversampling Technique (SMOTE)

The SMOTE technique [9] was proposed to combat the disavantages of the simple over-sampling and undersampling techniques. The SMOTE technique creates artificial data based on the feature space (rather than the data space) similarities from the minority modules. The SMOTE technique starts with a set of minority modules (i.e., defective modules). For each of the minority defective modules of the training datasets, SMOTE performs the following steps:

*(Step 1)* Calculate the $k$-nearest neighbors.
*(Step 2)* Select $N$ majority clean modules based on the smallest magnitude of the euclidean distances that are obtained from the $k$-nearest neighbors.

Finally, SMOTE combines the synthetic oversampling of the minority defective modules with the undersampling the majority clean modules.

### 3.4 Boostrap Random Over-Sampling Examples Technique (ROSE)

The ROSE technique [42] uses a smoothed-bootstrapping approach to draw artificial samples from the feature space neighbourhood around the minority class [17]. ROSE combines oversampling and undersampling by generating an augmented sample of the data (especially belonging to the rare class). The ROSE technique is made up of three steps:

*(Step 1)* Resampling the data of the majority class using a bootstrap resampling technique to remove modules of the majority class to a defective ratio of 50%. (undersampling)
*(Step 2)* Resampling the data of the minority class using a bootstrap resampling technique to repeat modules of the minority class to a defective ratio of 50%. (oversampling)

*(Step 3)* Generating a new synthetic data in its neighborhood, where the shape of the neighborhood is determined by the `rose.real()` R function that is provided by the ROSE R package.

These three steps are repeated for each training sample in order to produce a new synthetic training sample of approximately equal size as the original dataset where the number of modules for both classes equally represent (i.e., a defective ratio of nearly 50%).

## 4 RELATED WORK & RESEARCH QUESTIONS

Defect prediction models may produce inaccurate predictions and interpretation when they are trained on imbalanced datasets (i.e., a dataset where the proportion of defective and clean modules is not equally represented). Prior research investigated the impact of class rebalancing techniques on the performance of defect prediction models. For example, Kamei *et al.* [34] investigate the impact of class rebalancing techniques on 2 proprietary datasets using 4 classification techniques (i.e., linear discriminant analysis (LDA), logistic regression analysis (LRA), neural network (NN), and classification tree (CT)) and 3 performance measures (i.e., Precision, Recall, and F-Measure). Riquelme *et al.* [63] investigate the impact of class rebalancing techniques on 4 open-source datasets (i.e., CM1, KC1, KC2, PC1) using 2 classification techniques (i.e., Naive Bayes and C4.5) and 1 performance measure (i.e., AUC). Wang *et al.* [90] investigate the impact of class rebalancing techniques on 5 open-source datasets (i.e., CM1, KC3, PC1, PC3, MW1) using 2 classification techniques (i.e., Naive Bayes and AdaBoost) and 5 performance measure (i.e., PD, PF, Balance, G-mean, AUC). Tan *et al.* [78] investigate the impact of class rebalancing techniques on 7 commercial and open-source datasets using 7 classification techniques (i.e., Naive Bayes, Instance-based learning, Boosting, KNN, and SVM) and 3 performance measure (i.e., Precision, Recall, and F-Measure). However, prior work focus on a limited number of datasets and performance measures, which limits the generalization (i.e., external validity) of conclusions (see Table 5).

Indeed, the conclusions of prior research are contradictory. For example, Kamei *et al.* [34] find that class rebalancing techniques improve the F-measure performance by 7.8%-22.4%. However, Riquelme *et al.* [63] argue that class rebalancing techniques do not improve the percentage of correctly classified modules (i.e., Accuracy), but they do improve the AUC measure. A recent meta-analysis of 42 primary defect prediction studies [44] also demonstrates that class imbalance is not considered harmful when the minority class is above 20%. Such inconsistent conclusions make it hard to derive practical guidelines when applying class rebalancing techniques when constructing defect prediction models. To address the inconsistent conclusions and generalization issue of prior work, we address the following research question.

TABLE 1: An overview comparison of our study with respect to prior work.

| Study | #Classification | Datasets | Performance Measures |
|---|---|---|---|
| Kamei *et al.* [34] | 4 | 2 | P, R, and F1 |
| Riquelme *et al.* [63] | 2 | 4 | AUC |
| Wang *et al.* [90] | 2 | 5 | PD, PF, Balance, G-mean, AUC |
| Tan *et al.* [78] | 7 | 7 | P, R, and F1 |
| Our study | 7 | 101 | 10 performance measures |

*(RQ2) How do class rebalancing techniques impact the performance of defect prediction models?*

In addtion to being used for predictions, prior research also uses defect prediction models to uncover past pitfalls that lead to defective modules. For example, Hassan [24] studies the impact of complexity of code changes on software quality. Shihab *et al.* [73] investigate the impact of code and process metrics on post-release defects. Bettenburg *et al.* [4] investigate the impact of social interactions on software quality. McIntosh *et al.* [46] investigate the impact of code review coverage and participation on softwar equality. Thongtanunam *et al.* [87] investigate the impact of code review ownership on software quality. Such an understanding of defect characteristics is essential to chart quality improvement plans.

Recently, Turhan [89] point out that class rebalancing techniques may lead to bias in the learned concepts (i.e., *concept drift*). Yet, no research investigates the impact of class rebalancing techniques on the interpretation of defect prediction models. Thus, we address the following research question.

*(RQ3) How do class rebalancing techniques impact the interpretation of defect prediction models?*

## 5 CASE STUDY DESIGN

In this section, we describe the design of our case study that we perform to address our research questions. Figure 3 provides an overview of the case study design that we apply to each studied dataset. We describe each step below.

### 5.1 Studied Datasets

In selecting the studied datasets, we identified two important criteria that needed to be satisfied:

- **Criterion 1—Datasets from different corpora and domains.** Our recent work [84] shows the tendency of researchers to reuse experimental components (e.g., datasets, metrics, and classifiers) can introduce a bias in the reported results. To extend the generality of our conclusions, we choose to train our defect prediction models using datasets from as many different corpora and domains as possible.
- **Criterion 2—Publicly-available defect datasets**. Recently, replicability concerns are raised in our SE and medical discipline. For example, Robles *et al.* [64] point out that over 38% of 171 software engineering studies do not use publicly-available datasets nor provide their studied datasets. Moreover, Ioannidis *et al.* [29] raise concerns that the majority of medical studies in the highest ranked journals like Nature Genetic are not replicable. To foster the replication of our experiments, we choose to train our defect prediction models using datasets that are hosted in publicly-available data repositories.

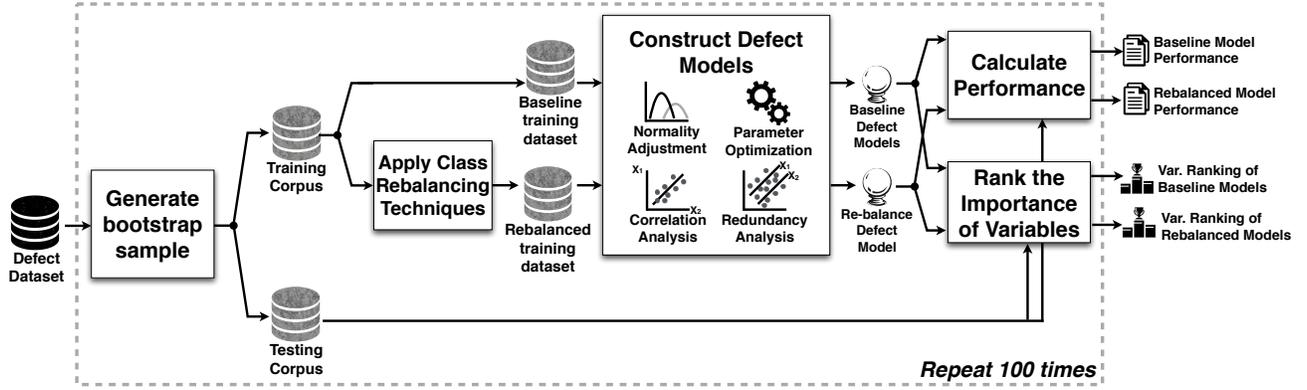

Fig. 3: An overview of our case study design.

To satisfy criterion 1 and 2, we opt to use the 101 publicly-available defect datasets that are described in Section 2. The 101 studied systems include proprietary and open source systems, with varying size, domain, and defective ratio.

## 5.2 Generate bootstrap samples

In order to ensure that the conclusions that we draw about our defect prediction models are robust, we use the out-of-sample bootstrap validation technique [85], which leverages aspects of statistical inference [16]. The out-of-sample bootstrap is made up of two steps:

*(Step 1)* A bootstrap sample of size $N$ is randomly drawn with replacement from an original dataset, which is also of size $N$.

*(Step 2)* A model is trained using the bootstrap sample and tested using the rows that do not appear in the bootstrap sample. On average, 36.8% of the rows will not appear in the bootstrap sample, since it is drawn with replacement [16].

The out-of-sample bootstrap process is repeated 100 times, and the average out-of-sample performance is reported as the performance estimate.

## 5.3 Apply Class Rebalancing Techniques

In order to investigate the impact of class rebalancing techniques, we apply the 4 studied class rebalancing techniques (as described in Section 3) only on the training datasets, while the testing data is not rebalanced. To apply the over-sampling technique, we use the implementation of the `upSample` function that is provided by the `caret` R package [40]. To apply the under-sampling technique, we use the implementation of the `downSample` function that is provided by the `caret` R package [40]. To apply the SMOTE technique, we use the implementation of the `SMOTE` function that is provided by the `DMwR` R package [88]. To apply the ROSE technique, we use the implementation of the `ROSE` function that is provided by the `ROSE` R package [42].

## 5.4 Construct Defect Models

There are a plethora of classification techniques that have been studied in defect prediction domain [19, 20, 41, 55, 72, 83]. Since it is impractical to study all of these techniques, we would like to select a manageable set of classification techniques for our study. In selecting the classification techniques for our study, we select to study only the top-ranked classification techniques, according to our recent analysis on the ranking of classification techniques for defect prediction models when automated parameter optimization is applied [86]. We choose 7 classification techniques that appear at the top-2 ranked classification techniques. We construct defect prediction models using 7 classification techniques, i.e., Random Forest (RF), Logistic Regression (LR), Naive Bayes (NB), neural network (AVNNet), C5.0 Boosting (C5.0), extreme gradient boosting (xGBTree), and gradient boosting method (GBM).

Random forest constructs multiple decision trees from bootstrap samples. Logistic regression measures the relationship between a categorical dependent variable and one or more independent variables. Naive bayes is a probability-based technique that assumes that all of the predictors are independent of each other. Neural network is used to estimate or approximate functions that can depend on a large number of inputs, and that are generally unknown. C5.0 Boosting, extreme gradient boosting (xGBTree), and the gradient boosting method (GBM) perform multiple iterations, each with different example weights, and makes predictions using classifier voting.

*Correlation Analysis.* Highly correlated independent variables may interfere with each other when a model is being interpreted. Indeed, our recent work [81, 84]

TABLE 2: The definitions and descriptions of our studied threshold-depending performance measures.

| Metrics | Definition | Description |
| --- | --- | --- |
| **Precision (P)** or Positive Predicted Values | $\frac{TP}{TP+FP}$ | A proportion of modules that are correctly classified as defective |
| **Recall (R)**, Probability of Detection (PD), True Positive Rate ($TP_{rate}$, Sensitivity) | $\frac{TP}{TP+FN}$ | A proportion of defective modules that are correctly classified |
| **F-Measure** | $2 \times \frac{P \times R}{P+R}$ | A harmonic mean of precision (P) and recall (R) |
| **Matthews Correlation Coefficient (MCC)** | $\frac{TP \times TN - FP \times FN}{\sqrt{(TP+FP)(TP+FN)(TN+FP)(TN+FN)}}$ | A balanced measure based on true and false positives and negatives |
| **G-mean** | $\sqrt{TP_{rate} \times TN_{rate}}$ | A geometric mean of a true positive rate and a true negative rate |
| **G-measure** | $\frac{2 \times PD \times (1 - FP_{rate})}{PD + (1 - FP_{rate})}$ | A harmonic mean of the probability of detection (PD) and a false positive rate ($FP_{rate}$) |
| **Accuracy** | $\frac{TP+TN}{TP+FN+FP+TN}$ | A proportion of correctly classified modules |

Note: TP = True Positive, TN = True Negative, FP = False Positive, FN = False Negative, $FP_{rate} = \frac{FP}{FP+TN}$

demonstrates that collinearity and multicollinearity issues can artificially inflate (or deflate) the impact of software metrics when interpreting defect prediction models. Jiarpakdee *et al.* [31] point out that 10%-67% of metrics of publicly-available defect datasets are redundant. Recently, Jiarpakdee *et al.* [30] point out that correlated metrics impact the ranking of the highest ranked metric of defect prediction models. Moreover, Jiarpakdee *et al.* [30] point out that removing correlated metrics improves the consistency of the highest ranked metric regardless of how a model is specified and negligibly impacts the performance and stability of defect models. Thus, we perform correlation and redundancy analyses prior to training our defect prediction models. We measure the correlation between explanatory variables using Spearman rank correlation tests ($\rho$). We then use a variable clustering analysis [67] to construct a hierarchical overview of the correlation and remove explanatory variables with a high correlation. We select $|\rho| = 0.7$ as a threshold for removing highly correlated variables [35]. We perform this analysis iteratively until all clusters of surviving variables have $|\rho| < 0.7$.

*Redundancy Analysis.* While correlation analysis reduces collinearity among our variables, it does not detect all of the *redundant variables*, i.e., variables that do not have a unique signal with respect to the other variables. Redundant variables will interfere with each other, distorting the modelled relationship between the explanatory variables and the outcome. Therefore, we remove redundant variables prior to constructing our defect prediction models. In order to detect redundant variables, we fit preliminary models that explain each variable using the other explanatory variables. We use the $R^2$ value of the preliminary models to measure how well each variable is explained by the others.

We use the implementation of redundancy analysis as provided by the `redun` function of the `rms` R package [22]. The variable that is most well-explained by the other variables is iteratively dropped until either: (1) no preliminary model achieves an $R^2$ above a cutoff threshold (for this paper, we use the default threshold of 0.9), or (2) removing a variable would make a previously dropped variable no longer explainable, i.e., its preliminary model will no longer achieve an $R^2$ exceeding the threshold.

*Parameter Settings.* Since the studied classification techniques have configurable parameter settings, we apply Caret parameter optimization [40] prior to constructing defect prediction models as suggested by Tantithamthavorn [83].

### 5.5 Calculate Performance

We apply the defect prediction models that we train using the training corpus to the untreated testing corpus (i.e., not rebalanced) in order to measure their performance. We use both threshold-independent and threshold-dependent performance measures to quantify the performance of our models. We describe the various performance measures that we used below.

#### 5.5.1 Threshold-Independent Performance Measures

First, we use the *Brier score* [6, 66] to measure the distance between the predicted probabilities and the outcome. The Brier score is calculated as $B = \frac{1}{N} \sum_{i=1}^{N} (f_t - o_t)^2$, where $f_t$ is the predicted probability, $o_t$ is the outcome for module $t$ encoded as 0 if module $t$ is clean and 1 if it is defective, and $N$ is the total number of modules. The Brier score ranges from 0 (best classifier performance) to 1 (worst classifier performance), where a Brier score of 0.25 is a random-guessing performance.

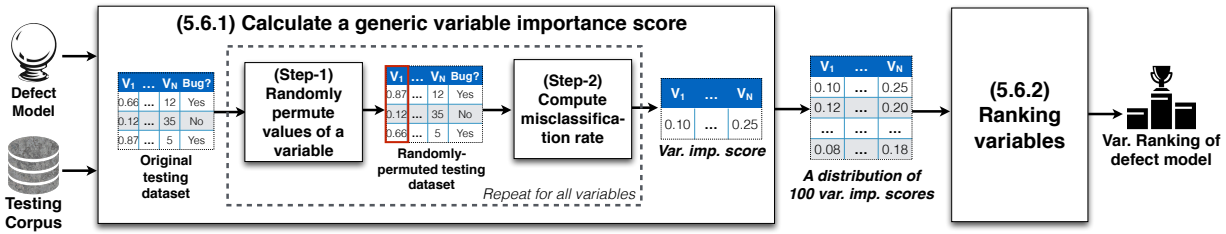

Fig. 4: An overview of our generic variable importance score calculation.

Second, we use the *calibration slope* to measure the direction and spread of the predicted probabilities [12, 15, 21, 23, 49, 75, 77]. The calibration slope is the slope of a logistic regression model that is trained using the predicted probabilities of our original defect prediction model to predict whether a module will be defective or not [12]. A calibration slope of 1 indicates the best classifier performance (i.e., the predicted probabilities are consistent with modules labels) and a calibration slope of 0 (or less) indicates the worst classifier performance (i.e., the predicted probabilities are inconsistent with module's labels)

Third, we use the *Area Under the receiver operator characteristic Curve (AUC)* to measure the discriminatory power of our models, as suggested by recent research [15, 21, 28, 41, 76, 77]. The AUC is a threshold-independent performance metric that measures a classifier's ability to discriminate between defective and clean modules (i.e., do the defective modules tend to have higher predicted probabilities than clean modules?). AUC is computed by measuring the area under the curve that plots the true positive rate against the false positive rate, while varying the threshold that is used to determine whether a file is classified as defective or not. Values of AUC range between 0 (worst performance), 0.5 (random guessing performance), and 1 (best performance). We use the `val.prob` function of the `rms` R package [22] to calculate the Brier score, calibration slope, and AUC.

### 5.5.2 Threshold-Dependent Performance Measures

In order to calculate the threshold-dependent performance measures, the probabilities are transformed into a binary classification (defective or clean) using a default threshold value of 0.5, i.e., if a module has a predicted probability above 0.5, it is considered defective; otherwise, the module is considered clean. Using the threshold of 0.5, we compute nine threshold-dependent performance measures. Table 2 provides the definitions and descriptions of our 7 threshold-dependent performance measures.

### 5.6 Rank the Importance of Variables

To identify the most important variables in our built models, we compute variable importance for each variable in our models. To do so, we develop a generic variable importance score that can be applied to any classifier [86]. Figure 4 provides an overview of the calculation of our variable importance measurement to generate ranks of important variables for each of the baseline and rebalanced models.

### 5.6.1 Generic Variable Importance Score

The calculation of our variable importance score consists of 2 steps for each variable.

*(Step 1)* For each testing dataset, we first randomly permute the values of that particular variable, producing a randomly-permuted dataset.

*(Step 2)* We then compute the difference in the misclassification rates of defect prediction models that are trained using the original-unpermuted and the randomly-permuted datasets. The larger the difference, the greater the importance of that variable.

We repeat the Steps 1 and 2 for each variable in order to produce a variable importance score for all variables. Since the experiment is repeated 100 times, each variable will have several variable importance scores (i.e., one score for each of the repetitions).

### 5.6.2 Ranking Variables

To study the impact of the studied variables on our models, we apply the improved Scott-Knott Effect Size Difference (ESD) test (v2.0) [79, 85, 86]. The Scott-Knott ESD test clusters variables according to statistically significant differences in their mean variable importance scores ($\alpha = 0.05$).

Unlike our earlier version of the Scott-Knott ESD test (v1.0) that post-processes the groups that are produced by the Scott-Knott test, the Scott-Knott ESD test (v2.0) checks the magnitude of the difference throughout the clustering process by merging pairs of statistically distinct groups that have a negligible Cohen's $d$ effect size difference for all of the treatments of those two groups. Cohen's $d$ effect size [10] is an effect size estimate based on the difference between the two means divided by the standard deviation of the two datasets ($d = \frac{\bar{x}_1 - \bar{x}_2}{s.d.}$). The magnitude is assessed using the thresholds that are provided by Cohen [11], i.e. $|d| < 0.2$ "negligible", $|d| < 0.5$ "small", $|d| < 0.8$ "medium", otherwise "large".

The Scott-Knott ESD test also overcomes the confounding factor of overlapping groups that are produced by other post-hoc tests [19, 50], such as Nemenyi's test [57], which were used in prior studies [41]. We use the implementation of the Scott-Knott ESD test (v2.0) that is provided by the `ScottKnottESD` R package [79].

Finally, we produce rankings of variables in the baseline models and rebalanced models. Thus, each variable has a rank for each type of model.

### 5.7 Statistical Analysis of the Experimental Settings

To better understand which of the experimental settings has the most impact on the performance of defect prediction models (i.e., RQ2 and RQ3), we build regression models to understand the relationship between experimental settings and outcome (e.g., performance difference). To study the importance of each configuration parameter, we perform an ANOVA analysis to examine the relative contribution (in terms of explanatory power) of each experimental settings to the regression model. Figure 5 shows an overview of our sensitivity analysis approach. We describe each step of our approach below.

**(Step-1) Construct Models for Experimental Settings.** We build regression models to explain the relationship that experimental settings have on the performance difference of defect prediction models. A regression model fits a line of the form $y = \beta_0 + \beta_1 x_1 + \beta_2 x_2 + ... + \beta_n x_n$ to the data, where $y$ is the dependent variable and each $x_i$ is an explanatory variable. We fit our regression models using the Ordinary Least Squares (OLS) technique using the `ols` function provided by the `rms` R package [22].

**(Step-2) Assessment of Model Stability.** We evaluate the fit of our models using the $Adjusted\ R^2$, which provides a measure of fit that penalizes the use of additional degrees of freedom. However, since the adjusted $R^2$ is measured using the same data that was used to train the model, it is inherently upwardly biased, i.e., "optimistic". We estimate the optimism of our models using the following bootstrap-derived approach [21].

First, we build a model from a bootstrap sample, i.e., a dataset sampled with replacement from the original dataset, which has the same population size as the original dataset. Then, the optimism is estimated using the difference of the adjusted $R^2$ of the bootstrap model when applied to the original dataset and the bootstrap sample. Finally, the calculation is repeated 1,000 times in order to calculate the average optimism. This average optimism is subtracted from the adjusted $R^2$ of the model fit on the original data to obtain the optimism-reduced adjusted $R^2$. The smaller the average optimism, the higher the stability of the original model fit.

**(Step-3) Estimate Power of Explanatory Variables.** We perform an ANOVA analysis to examine the relative contribution (in terms of explanative power) of each experimental settings to the regression models using the Wald $\chi^2$ maximum likelihood (a.k.a., "chunk") test. The larger the Wald $\chi^2$ value, the larger the impact that a particular explanatory variable has on the response [21]. Finally, we present both the raw Wald $\chi^2$ values, and its bootstrap 95 percentile confidence interval.

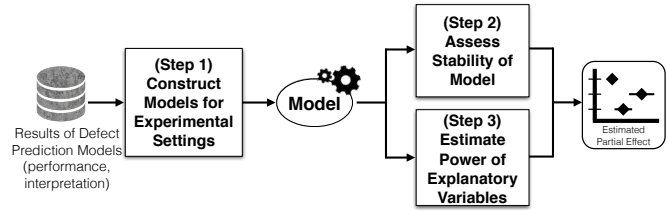

Fig. 5: An overview approach of statistical analysis of the experimental settings.

## 6 CASE STUDY RESULTS

In this section, we present the results of our case study with respect to the following two research questions.

**(RQ2) How do class rebalancing techniques impact the performance of defect prediction models?**

**Approach.** To address RQ2, we start with the performance distribution of defect prediction models that are trained using original (i.e., unbalanced) and re-balanced datasets. For each class rebalancing technique, we compute the difference in the performance of classifiers that are trained using using original and re-balanced datasets. We then use boxplots to present the distribution of the absolute performance difference for each of the 10 commonly-used performance measures.

**Results.** Figure 6 shows the absolute performance difference when applying class rebalancing techniques to defect prediction models for each of the 10 commonly-used performance measures. Figure 6 shows that the ROSE technique produces the least stable conclusions when applied to defect prediction models. Indeed, we observe that the ROSE technique has both positive and negative impact on the Recall, F-measure, G-measure, Gmean, and Slope of defect prediction models, suggesting that the ROSE technique should be avoid in future defect prediction studies. In the remainder of this section, we only focus on the under-sampling, over-sampling, and SMOTE techniques.

Below, we structure the following findings with respect to the performance measures that (1) are not impacted; (2) are improved; and (3) are decreased by class rebalancing techniques.

**The AUC measure is less sensitive to class rebalancing techniques, unlike common findings in machine learning domain that AUC performance is substantially improved.** Looking at Figure 6, when applying over-sampling, under-sampling, and SMOTE techniques, we observe that the absolute differences of AUC measure (i.e., `min-max`) vary from -4 to 7 percentage points. The absolute differences that we observe are relatively smaller than common findings in the machine learning domain. For C5.0 classification technique, we observe that the maximum AUC improvement is by up to 5 percentage points when applying SMOTE technique to defect prediction models. We note that C5.0 classification technique is an improvement of C4.5 classifier

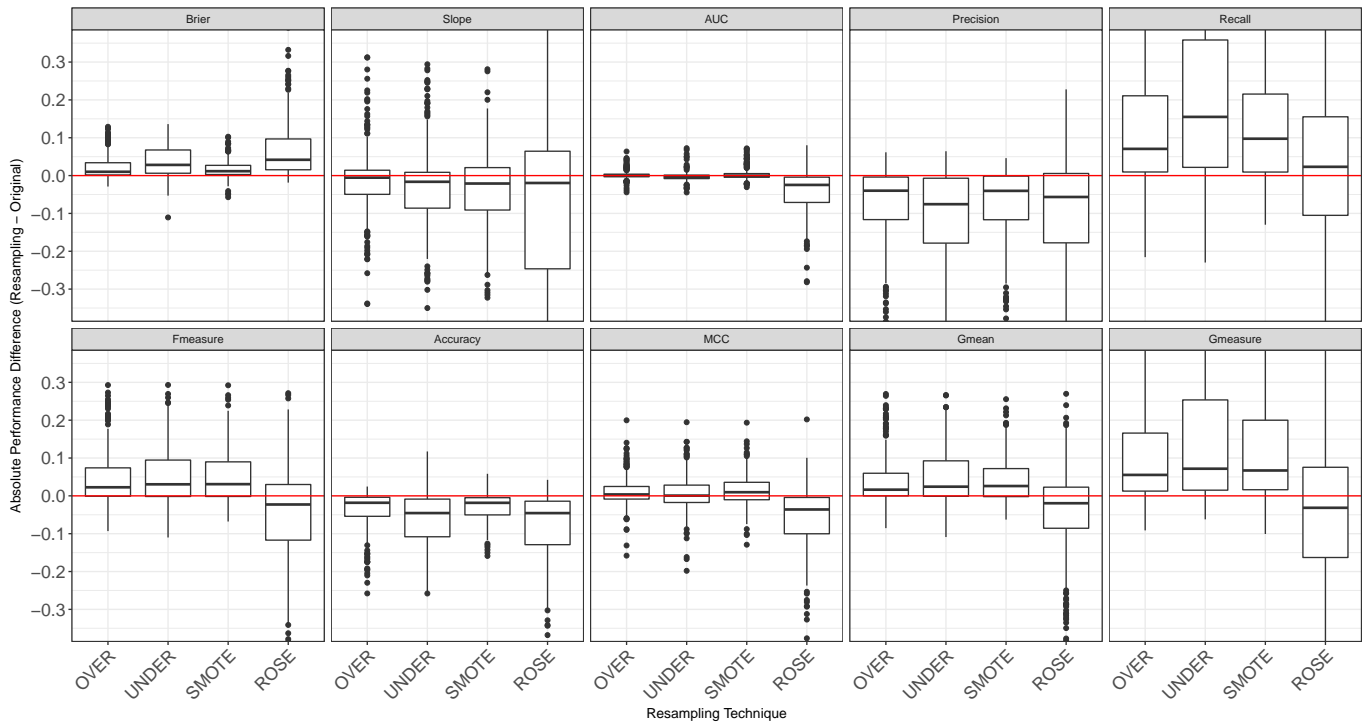

Fig. 6: The absolute performance difference when applying class rebalancing techniques to defect prediction models for each of the 10 commonly-used performance measures. A red line indicates a performance difference of zero (i.e., no improvement).

in terms of speed and memory usage while sharing similar algorithms to C4.5 classifiers [39]. This observation contradicts the conclusions of Chawla [8], who found that C4.5 classifiers tends to achieve a 40% of AUC improvement when applying SMOTE to machine learning datasets. Our contradictory observation shows that domain-specifics play an important role—*common findings about top-performing class rebalancing techniques in machine learning domain might not always hold true for software engineering domain.*

Moreover, we also observe similar trends with the MCC measure which is less sensitive to class rebalancing techniques. We find that the distributions of the absolute performance difference of AUC and MCC are centered at zero. For example, when applying over-sampling, under-sampling, and SMOTE techniques, we find that the absolute differences of the AUC measure for 75 percentage points of the defect prediction models (i.e., 1st-3rd quantiles) vary from -0.8 to 0.5 percentage points. The absolute differences of the MCC measure for 75 percentage points of the defect prediction models (i.e., 1st-3rd quantiles) vary from -1.7 to 3.6 percentage points. The distributions that are centered at zero indicate that the AUC and MCC measures are not impacted positively nor negatively by class rebalancing techniques for defect prediction models. The AUC measure is insensitive to class rebalancing techniques since it considers all probability thresholds for determining a module is defective or clean. On the other hand, the MCC measure is insensitive to class rebalancing techniques since it considers all aspects of the confusion metrics (i.e., true and false positives and negatives).

In contrast to the AUC measure which is rarely impacted, the other threshold-independent measures like the Brier and Slope measures are sensitive when applied class rebalancing techniques. The sensitivity of the Brier and Slope measures has to do with the computation of the Brier and Slope measures. Such computation relies heavily on the predicted probabilities (see Section 5.5.1). The Brier measure uses the predicted probabilities to compute the distance between the predicted probabilities and the outcome. The Slope measure uses the predicted probabilities to compute their directions and spreads.

**Class rebalancing techniques substantially improve the performance of defect prediction models by up to 69 percentage points for Recall, 60 percentage points for G-measure, for 27 percentage points of F-measure.** We find that the proportion of defective modules that are correctly classified (i.e., Recall) improves by up to 69 percentage points when applying the over-sampling technique to the `mylyn` dataset when constructing a logistic regression classifier. Moreover, we find that the F-measure improves by up to 27 percentage points when applying under-sampling technique to the `prop-5` dataset prior to constructing a logistic regression classifier. These results indicate that class rebalancing techniques tend to have a positive impact on the Recall, G-measure, and F-measure when they are applied to

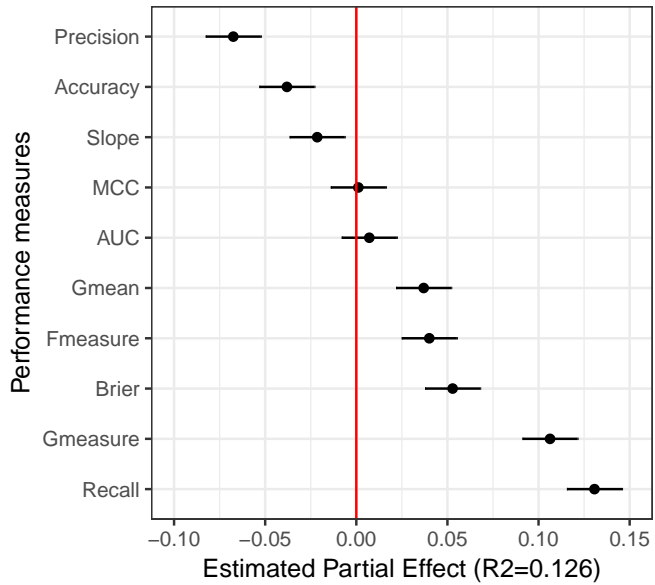

Fig. 7: Estimated partial effect plot of the relationship between performance measures and the magnitude of the performance difference with the 95% confidence interval.

defect prediction models.

**On the other hand, class rebalancing techniques decrease the performance of defect prediction models by up to 57 percentage points for Precision, 73 percentage points for Accuracy.** Interestingly, while class rebalancing techniques substantially increase Recall, they decrease Precision. We find that the decrease in the Precision measure has to do with an increased number of false positive (FP) modules (i.e., the number of clean modules that are misclassified), suggesting that the improvement of the F-measure performance has to do with the improvement of the proportion of defective modules that are correctly classified (i.e., Recall).

**Statistical Analysis of the Performance Measures**. To statistically (1) validate if the distributions of the performance measures are statistically different; and (2) investigate which performance measures have the largest and smallest impact, we construct a one-way ANOVA model of the distributions of the performance measures. The one-way ANOVA is a hypothesis test in which a single categorical variable or a single factor (i.e., performance measures) is considered when comparing the mean distributions of the performance values for all of the 10 studied performance measures. The one-way ANOVA model confirms that there is a significant difference in the means among the performance measures with a significant level of 0.05 (i.e., $p$-value $< 0.05$). We then plot the estimated partial effect of the performance differences with the 95% confidence interval (see Figure 7). The x-axis describes the effect of the performance differences, while the y-axis describes the performance measures. The effect values indicate the positive and negative magnitude of the performance differences, while an effect value of zero indicates that class rebalancing techniques have no impact to a particular performance measure. The estimated partial effect plot of Figure 7 confirms that AUC and MCC are insensitive to class rebalancing techniques. Moreover, Figure 7 also confirms that class rebalancing techniques yield the largest positive impact on Recall and the largest negative impact on Precision.

> *The AUC measure is not impacted by class rebalancing techniques for defect prediction models, unlike common findings in machine learning literature where the AUC performance is substantially improved with the applications of class rebalancing techniques. Class rebalancing techniques impact Recall the most positively and impact Precision the most negatively.*

**Statistical Analysis of the Experimental Settings**. To better understand the experimental settings where class rebalancing techniques yield the largest positive impact on Recall, we construct a regression model to understand the relationship between the experimental settings (i.e., EPV, defective ratio, classification techniques, class rebalancing techniques, and metric family) and the performance difference of the Recall measure using the high-level approach of Section 5.7. *EPV (Event-Per-Variables)*—a measure to evaluate the risk of overfitting—is the ratio of events to the number of independent variables used to train a model. Formally,

$$EPV = \frac{\#events \text{ (e.g., \#defective modules)}}{\#variables} \quad (1)$$

where the event is the number of occurrences of the least frequently occurring class of the dependent variable (e.g., the numbers of defective modules), and the variables is the number of independent variables used to train the model (i.e., the number of software metrics) [85]. Recently, Tantithamthavorn *et al.* [85] demonstrated that models that are trained using datasets where the EPV is low (i.e., too few events are available relative to the number of independent variables) are especially susceptible to overfitting (i.e., being fit too closely to the training data). We perform a statistical analysis only for the Recall measure, since we find that class rebalancing techniques yield the largest positive impact on Recall. We included the EPV measure in our statistical models in order to control for a confounding factor of dataset characteristics. Since we experiment on 101 defect datasets where each dataset has different set of metrics, it is possible that some metrics are robust or sensitive to class rebalancing techniques. To statistically verify this assumption while controlling for other factors, we also include a Metric Family (i.e., D'Ambros, CK, Eclipse, Kim&Wu, and McCabe) into our statistical models. Table 3 shows the statistics of the regression model.

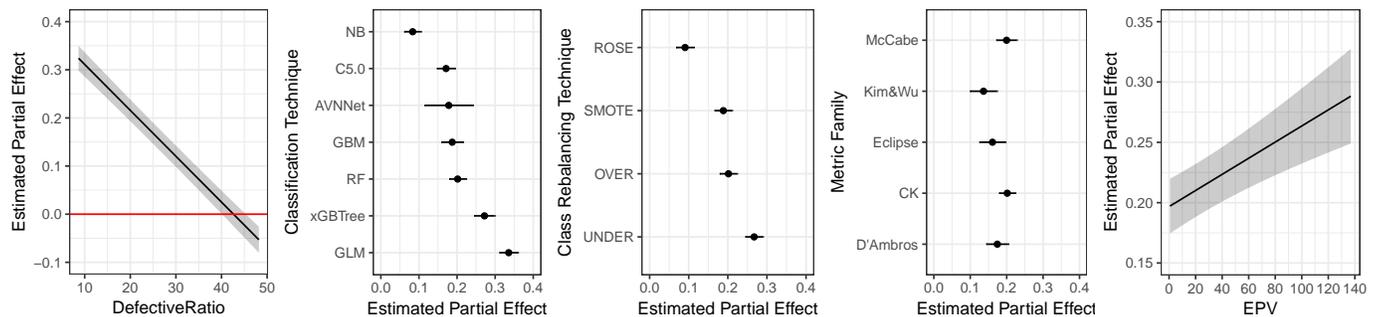

Fig. 8: Estimated partial effect plots of the relationship between the experimental settings (i.e., Defective Ratio, Classification Technique, Class Rebalancing Technique, Metric Family, and EPV) and the performance difference of the Recall measure. The grey areas and error bars indicate the 95% confidence interval. The effect values indicate the positive and negative magnitude of the performance differences, while an effect value of zero indicates no performance differences.

TABLE 3: Statistics of the regression model of the relationship between the experimental factors and the performance difference of the Recall measure.

|  | | Factor Analysis | |
|---|---|---|---|
| Adjusted $R^2$ | | 0.62 | |
| Optimism-reduced adjusted $R^2$ | | 0.60 | |
| Total Wald $\chi^2$ | | 1,350 | |
|  | D.F. | $\chi^2$ | $p$-value |
| Defective Ratio | 1 | 46% | *** |
| Classification Technique | 6 | 33% | *** |
| Class Rebalancing Technique | 3 | 19% | *** |
| Metric Family | 4 | 1% | *** |
| EPV | 1 | 1% | *** |

Statistical significance of explanatory power according to Wald $\chi^2$ likelihood ratio test: $\circ$ $p \geq 0.05$; * $p < 0.05$; ** $p < 0.01$; *** $p < 0.001$

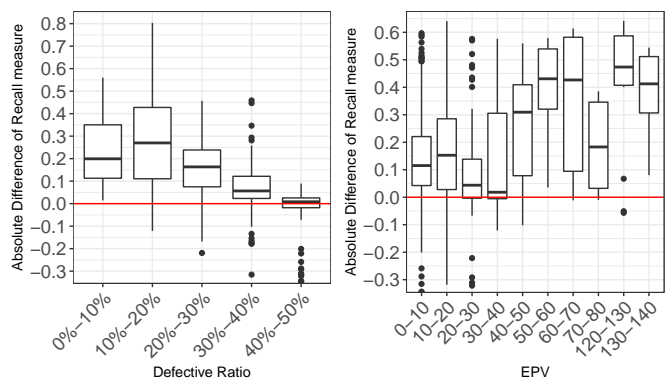

Fig. 9: The performance difference for the Recall measure for each range of defective and EPV ratios.

**The impact of class rebalancing techniques on the performance of defect prediction models relies heavily on the defective ratio of the defect datasets.** Table 3 shows that the *DefectiveRatio* is the most influential experimental factor that impacts the performance of defect prediction models. Figure 8 also shows that the impact of experimental settings has a large impact on the performance of defect prediction models when applying class rebalancing techniques. On the other hand, Table 3 shows that Metric Family and EPV has little impact on the performance of defect prediction models when applying class rebalancing techniques. We discuss the impact of experimental settings on the performance difference of class rebalancing techniques for defect prediction models below.

**For the Recall performance measure, class rebalancing techniques yield the largest benefits when they are applied to highly-imbalanced defect datasets with a defective ratio below 20%.** The estimated partial effect plot of Figure 8 for defective ratios shows a negative relationship between the defective ratios of defect datasets and the performance difference of defect prediction mod-

els. Thus, we plot the performance difference of the Recall measure for each range of the defective ratios in Figure 9. We find that class rebalancing techniques tend to yield the largest performance improvement for defect prediction models when they are applied to datasets with a defective ratio below 20%. Specifically, for highly-imbalanced defect datasets (i.e., a defective ratio below 10%), we observe that class rebalancing techniques consistently improve the Recall of defect prediction models. On the other hand, for nearly-balanced defect datasets (i.e., a defective ratio between 40-50%), we find that class rebalancing techniques have little impact on the performance improvement.

**Class rebalancing techniques yield the largest benefits when they are applied to defect datasets with an EPV ratio higher than 40.** The estimated partial effect plot of Figure 8 for EPV ratios shows a positive relationship between the EPV ratios of defect datasets and the impact on the performance of defect prediction models. Thus, we plot the performance difference of the Recall measure for each range of the EPV ratios in Figure 9. We observe that class rebalancing techniques tend to yield the largest performance improvement for

defect prediction models when they are applied to defect datasets with an EPV ratio higher than 40. Our finding is consistent with Blagus *et al.* [5]'s findings which points out that SMOTE does not perform well with high-dimensionality data.

**Logistic regression is the most sensitive classifier to imbalanced defect datasets, while more advanced classification techniques like neural networks and random forest tend to be less sensitive.** Figure 8, the estimated partial effect plot for classification techniques, shows that the impact of class rebalancing techniques on the performance of defect prediction models varies between the studied classification techniques. We find that logistic regression classifiers tend to yield the largest benefit, while naive bayes classifiers tend to yield the smallest benefit on the Recall performance measure. Thus, class rebalancing techniques should be applied to future defect prediction studies which making use of logistic regression.

**The under-sampling technique improves Recall measure the most.** Figure 8, the estimated partial effect plot for classification techniques, shows a positive rationship between the class rebalancing techniques and the performance of defect prediction models. We find that the under-sampling technique tends to improve Recall measure the most, while the ROSE technique tends to improve Recall measure the least. Similarly, Figure 6 shows that the Recall improvement is, on average, 18 percentage points for under-sampling technique, 10 percentage points for the SMOTE technique, 7 percentage points for the over-sampling technique, and 3 percentage points for the ROSE technique, indicating that the under-sampling technique should be used when the main objective of defect prediction model is the proportion of defective modules that are correctly classified (i.e., Recall).

> The impact of class rebalancing techniques on the performance of defect prediction models depends on experimental settings. Class rebalancing techniques yield the largest performance improvement for defect prediction models when applying the under-sampling technique to logistic regression models using defect datasets that are highly-imbalanced with an EPV ratio higher than 40.

### (RQ3) How do class rebalancing techniques impact the interpretation of defect prediction models?

**Approach**. To address RQ3, we start with the variable ranking of the 7 studied classification techniques on each of the 101 studied datasets for both classifiers that are trained with the original and rebalanced datasets. For each classification technique, we compute the difference in the ranks of the variables that appear in the top-three ranks of the classifiers that are trained using the original and rebalanced datasets. For example, if a variable $v$ appears in the top rank in both the original and rebalanced models, then the variable would have a rank difference of 0. However, if $v$ appears in the third rank in the rebalanced model, then the rank difference of $v$ would be -2.

TABLE 4: Statistics of the regression model of the relationship between the experimental factors and the percentage of the most important variables appearing in the same rank.

|  | Factor Analysis | | |
|---|---|---|---|
| Adjusted $R^2$ | 0.15 | | |
| Optimism-reduced adjusted $R^2$ | 0.14 | | |
| Total Wald $\chi^2$ | 406 | | |
|  | D.F. | $\chi^2$ | *p*-value |
| Classification Technique | 6 | 79% | *** |
| Metric Family | 4 | 10% | *** |
| Class Rebalancing Technique | 3 | 5% | *** |
| Defective Ratio | 1 | 4% | *** |
| EPV | 1 | 1% | ○ |

Statistical significance of explanatory power according to Wald $\chi^2$ likelihood ratio test: ○ $p \geq 0.05$; * $p < 0.05$; ** $p < 0.01$; *** $p < 0.001$

**Results**. **Class rebalancing techniques have a large impact on the interpretation of defect prediction models that are produced by popularly-used classification techniques like random forest, logistic regression, and neural network.** Figure 10 shows that as little as 23%-34%, 55%-62%, and 68%-71% of the top variables in the top importance rank of the rebalanced models appear in the top importance rank of the baseline models for neural network, logistic regression, and random forests classifiers, respectively. In other words, as much as 77%-66%, 45%-38%, and 32%-29% of the top variables in the top importance rank of the rebalanced models do not appear in the top importance rank of the original models. Moreover, the variables in the second and third ranks are even more unstable. *This is the first empirical evidence that confirms the suspicious of Turhan [89] who point out that class rebalancing techniques shift the learned concepts (i.e., biasing the interpretation of defect prediction models)*, suggesting that class rebalancing techniques should be avoided in future defect prediction studies, especially, when deriving knowledge and understandings from these models.

> Class rebalancing techniques shift the learned concepts (i.e., biasing the interpretation of defect prediction models). We find that as little as 23%-34%, 55%-62%, and 68%-71% of the top variables in the top importance rank of the rebalanced models appear in the top importance rank of the baseline models for neural network, logistic regression, and random forest classifiers, respectively.

**Statistical Analysis on the Experimental Settings**. To better understand the experimental setup where class rebalancing techniques have the largest and smallest impact on the interpretation of defect prediction models, we construct a regression model to understand the relationship between the experimental factors (i.e., EPV, defective ratio, classification techniques, class rebalancing techniques, and metric family) and the percentage of the most important variables appearing in the same rank using the high-level approach of Section 5.7. Table 4 shows the statistics of the regression model. Figure 7 shows the estimated partial effect of the relationship between performance measures and the percentage of the most important variables appearing in the same rank.

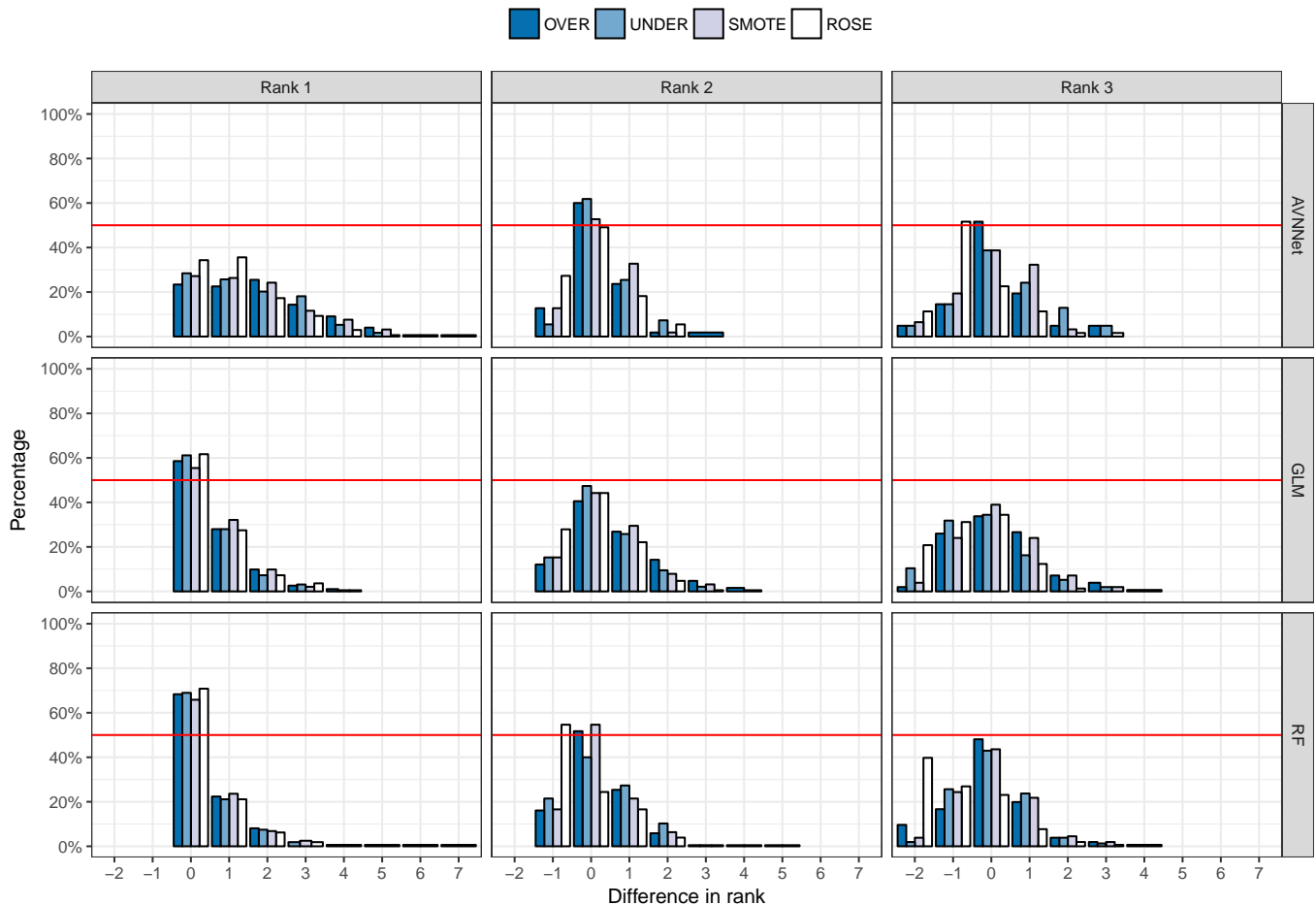

Fig. 10: The difference in the ranks for the variables according to their variable importance scores among the defect prediction models that are trained using original (i.e., baseline) and re-balanced datasets. The bars indicate the percentage of variables that appear in that rank in the re-balanced model while also appearing in that rank in the baseline model. The higher the percentage is, the least stable the interpretation of defect prediction models is.

**The impact of class rebalancing techniques on the interpretation of defect prediction models relies heavily on the used classification techniques.** Table 4 shows that *classification technique* is the most influential experimental factor on the impact of class rebalancing techniques on the interpretation of defect prediction models. Figure 11 confirms that neural network classifiers are the most sensitive classification techniques when applying class rebalancing techniques, suggesting that neural network classification techniques should be avoided when interpreting insights from defect prediction models.

> *The impact of class rebalancing techniques on the interpretation of defect prediction models relies heavily on the used classification techniques, suggesting that researchers and practitioners should avoid class rebalancing techniques when deriving knowledge and understandings from defect prediction models.*

## 7 PRACTICAL GUIDELINES

Our experimental results indicate that class rebalancing techniques have little impact on the AUC performance measure and the interpretation of defect prediction models that are trained using popularly-used classification techniques like logistic regression and random forest. Table 5 summarizes a comparison of prior findings in the literature with our findings. In this section, we offer practical guidelines for future defect prediction studies:

**(1) Class rebalancing techniques are beneficial when quality assurance teams wish to increase the completeness of identifying software defects (i.e., Recall)**, since Figure 7 shows that class rebalancing

TABLE 5: A comparison of prior findings in the literature to our findings.

| Paper | Their finding | Our finding |
|---|---|---|
| Riquelme *et al.* [63] | Class rebalancing techniques do not improve the percentage of correctly classified modules (i.e., Accuracy), but they do improve the AUC measure for 4 NASA datasets. | Class rebalancing techniques do improve the percentage of correctly classified modules (i.e., Accuracy), but they do not improve the AUC measure for 101 defect datasets. |
| Chawla [8] | SMOTE improves the AUC measure by up to 40% for machine learning datasets. | The AUC measure is less sensitive to class rebalancing techniques for 101 defect datasets. |
| Wang *et al.* [90] | An advanced class rebalancing technique (i.e., AdaBoost.NC) yeilds similar AUC performance when compared to random forest for 10 NASA defect datasets. | The AUC measure is less sensitive to class rebalancing techniques for 101 defect datasets. |
| Tan *et al.* [78] | Class rebalancing techniques improve the Precision measure but decrease the Recall measure for change classification models. | Class rebalancing techniques decrease the Precision measure but improve the Recall measure for defect classification models. |
| Kamei *et al.* [34] | Class rebalancing techniques improve the Recall and F-measures but decrease the Precision measure. | Class rebalancing techniques improve the Recall and F-measures but decrease the Precision measure. |

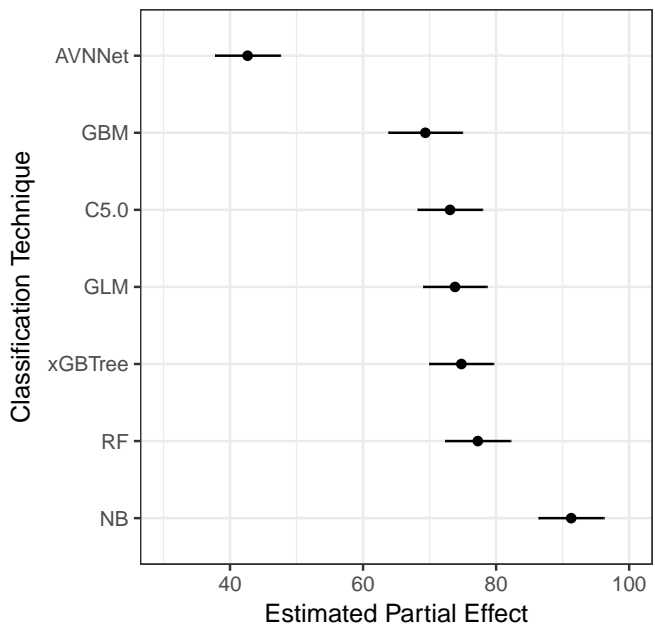

Fig. 11: Estimated partial effect plot of the relationship between the studied classification techniques and the percentage of the most important variables appearing in the same rank. The error bars indicate the 95% confidence interval. The figure shows that neural network is the most sensitive classification technique, while naive bayes is the least sensitive classification technique to class rebalancing techniques.

techniques substantially improve the proportion of defective modules that are correctly classified (i.e., Recall measure). Specifically, Figure 8 also shows that defect prediction models yield the largest improvement in Recall when applying the under-sampling technique to logistic regression models using defect datasets that are highly-imbalanced with an EPV ratio higher than 40. Nevertheless, when applying class rebalancing techniques, the improvement of the Recall measure has to be sacrificed with a decrease in the Precision measure.

(2) **Class rebalancing techniques should be avoided when deriving knowledge and understandings from defect prediction models to initiate quality improvement plans**, since Figure 11 shows that commonly-used classification techniques like logistic regression, random forest, and neural networks are sensitive to class rebalancing techniques. Moreover, our statistical model (Table 4) of the experimental factors analysis confirms that class rebalancing techniques have a large impact on the interpretation of defect prediction models.

(3) **AUC should be used as a standard measure for comparing defect prediction models**, since Figure 6 and 7 show that AUC is insensitive to class rebalancing techniques and imbalanced datasets that are commonly present in the defect prediction domain. Class rebalancing techniques primarily impact the probability scores that are produced by defect prediction models. Thus, class rebalancing techniques substantially impact the performance measures that rely heavily on a probability threshold (e.g., Precision, Recall, F-measure, and Brier score). On the other hand, the AUC measure is insensitive to class rebalancing techniques since it considers all probability thresholds for determining a module is defective or clean. Moreover, the MCC measure is insensitive to class rebalancing techniques since it considers all aspects of the confusion metrics (i.e., true and false positives and negatives).

# 8 THREATS TO VALIDITY

Like any empirical study design, experimental design settings may impact the results of our study [80].

Below, we discuss threats that may impact the results of our study.

## 8.1 External Validity

We studied a limited number of proprietary and open-source software systems. Thus, our results may not generalize to all software systems. However, to the best of our knowledge, this study is among the largest empirical study on the impact of class rebalancing techniques for defect prediction models — our conclusions are drawn from the 101 publicly-available defect datasets.

The conclusions of our case study rely on one defect prediction scenario (i.e., within-project defect prediction models). However, there are a variety of defect prediction scenarios in the literature (e.g., cross-project defect prediction [95], just-in-time defect prediction [32], heterogenous defect prediction [56]). Therefore, the practical guidelines may differ when applying class rebalancing techniques to other scenarios. Thus, future research should revisit our study in other scenarios of defect prediction models.

Recent studies [33, 59–61] recommend the consideration of developer effort when evaluating the performance of defect prediction models. For example, Kamei *et al.* [60] suggest to evaluate defect prediction models using the Area Under the Cost Effectiveness Curve (AUCEC). While we studied a large number of performance measures, i.e., 7 threshold-dependent and 3 threshold-independent measures, our results may not generalize to other performance measures (e.g., AUCEC). Since the AUCEC measure is not currently compatible with the Caret implementation [40]. Caret measures the model performance based on a `summaryFunction` function that only takes the observed and predicted values. Hence, we are unable to compute the AUCEC measure. Nonetheless, other performance measures can be explored in future work. We provide a detailed methodology for others who would like to re-examine our findings using unexplored performance measures.

### 8.2 Internal Validity

Recent work pointed out that class rebalancing techniques suffer from creating an artificial bias towards minority class [43]. Thus, Friedman *et al.* [18] suggested to use advanced classification techniques (e.g., penalized classification) to address the class imbalance problem for defect prediction models without applying a class rebalancing technique. Such penalization classifcation imposes an additional cost on the models for making classification mistakes on the minority class during training, while enabling the models to pay more attention to the minority class. To assess if a penalized classification technique addresses the class imbalance problem for defect prediction models, we built penalized logistic regression models for the 101 studied defect datasets using the `glmnet` function that is provided by the `glmnet` R package [25]. We then compared the AUC distributions with the other seven classification techniques when both class rebalancing techniques are applied and not applied. We find that the penalized logistic regression models have little improvement on the AUC performance. Figure 12 shows the AUC distributions of the 101 defect datasets for each of the studied seven classification technique, the four class rebalancing techniques, and the penalized logistic regression technique. The results of

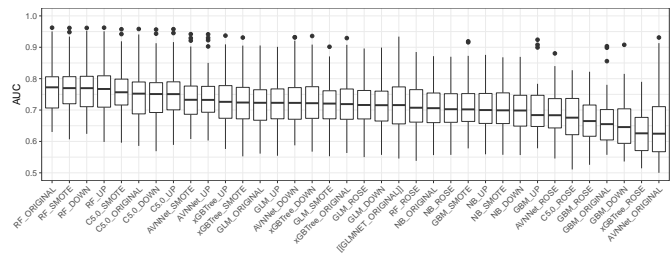

Fig. 12: The AUC distributions of the 101 defect datasets for each of the studied seven classification technique, the four class rebalancing techniques, and the penalized logistic regression technique.

Figure 12 confirms that building random forest models without applying class rebalancing techniques tends to be the top-performing classifier for the AUC performance. Nevertheless, other class rebalancing techniques should be explored in future work [27]. We provide a detailed methodology for others who would like to re-examine our findings using unexplored class rebalancing techniques and advanced classification techniques.

### 8.3 Construct Validity

Plenty of prior work show that the parameters of classification techniques have an impact on the performance of defect prediction models [83, 86]. Similarly, the SMOTE class rebalancing technique has a configurable parameter that need to be specified. Similar to prior studies in software engineering [3, 69, 92], the results of our study rely on one default parameter setting (i.e., $k = 5$). However, a recent study by Agrawal and Menzies [1] pointed out that different SMOTE parameter settings might provide different results. To ensure that our conclusions are not sensitive to SMOTE parameter settings, we repeat the experiment with $k$=14, since the optimal $k$ parameter as found by Agrawal and Menzies ranges from 11 to 14. Figure 13 shows the AUC distributions of defect prediction models when the SMOTE rebalancing technique is applied with 2 parameter settings, i.e., $k$=5, and $k$=14. The Mann-Whitney U test confirms that the two distributions are not statistically significant ($p$-value=0.9957). Thus, we suspect that the $k$ parameter of the SMOTE rebalancing technique does not alter the conclusions of our paper.

## 9 CONCLUSIONS & FUTURE WORK

In this paper, we set out to investigate the impact of 4 popularly-used class rebalancing techniques, i.e., over-sampling, under-sampling, SMOTE, and ROSE techniques, on the performance and the interpretation of defect prediction models. We train our defect prediction models using 7 classification techniques and evaluate the performance of defect prediction models using 10 commonly-used performance measures. To better understand in which experimental settings class rebalancing techniques are beneficial for defect prediction models, we also construct statistical models to study the relationship between the experimental settings and the performance and interpretation of defect prediction models. Through a large-scale empirical study of 101 publicly-available defect datasets that span across open-source and proprietary systems that are collected from 5 different corpus, we record the following observations:

- As little as 8% of defect datasets have a defective ratio between 45%-55%, suggesting that most defect datasets are unbalanced.
- The AUC measure is not impacted by class rebalancing techniques for defect prediction models, unlike common findings in machine learning literature where the AUC performance is substantially improved with the applications of class rebalancing techniques. Class rebalancing techniques impact Recall the most positively and impact Precision the most negatively.
- Class rebalancing techniques yield the largest performance improvement for defect prediction models when applying the under-sampling rebalancing technique to logistic regression models using highly-imbalanced and low-dimensionality defect datasets.
- Unfortunately, class rebalancing techniques shift the learned concepts (i.e., biasing the interpretation of defect prediction models). We find that as little as 23%-34%, 55%-62%, and 68%-71% of the top variables in the top importance rank of the re-balanced models appear in the top importance rank of the baseline models for neural network, logistic regression, and random forest classifiers, respectively.
- The impact of class rebalancing techniques on the interpretation of defect prediction models relies heavily on the used classification techniques, suggesting that researchers and practitioners should avoid class rebalancing techniques when deriving knowledge and understandings from defect prediction models.

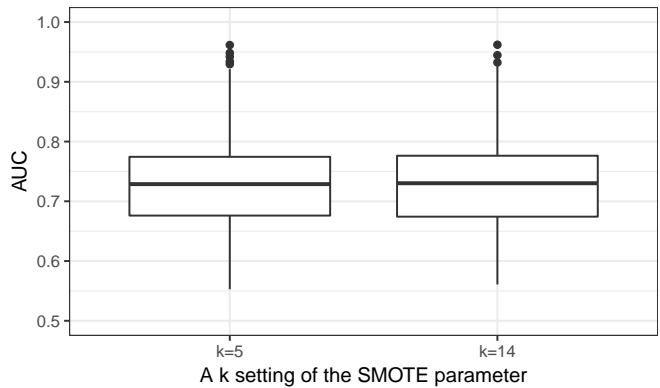

Fig. 13: The AUC distributions of defect prediction models when the SMOTE rebalancing technique is applied with 2 parameter settings, i.e., $k$=5, and $k$=14.

Based on our findings, we make the following suggestions for researchers and practitioners:

1) Class rebalancing techniques are beneficial when quality assurance teams wish to increase the completeness of identifying software defects (i.e., Recall)
2) Class rebalancing techniques should be avoided when deriving knowledge and understandings from defect prediction models to initiate quality improvement plans.
3) AUC should be used as a standard measure for comparing defect prediction models.

## ACKNOWLEDGMENTS

This study would not have been possible without High Performance Computing (HPC) systems provided by the Compute Canada[1] and the Centre for Advanced Computing at Queen's University.[2] This work was supported by the Grant-in-Aid for JSPS Fellows (No. 16J03360), and the Natural Sciences and Engineering Research Council of Canada (NSERC).